\documentclass[prc,floatfix,groupedaddress,nofootinbib,showpacs,preprintnumbers,
amsmath,amssymb,amsfonts,superscriptaddress,widetable] {revtex4-1}
\usepackage{bm}
\usepackage{mathrsfs}
\usepackage{amssymb}
\usepackage{amsmath}
\usepackage{graphicx}
\usepackage{array}
\usepackage[usenames,x11names]{xcolor}

\begin{document}
\title{Pulsar Glitches: The Crust may be Enough}
\author{J. Piekarewicz}\email{jpiekarewicz@fsu.edu}
\affiliation{Department of Physics, Florida State University,
               Tallahassee, FL 32306, USA}
\author{F.~J. Fattoyev}\email{farrooh.fattoyev@tamuc.edu}
\affiliation{Department of Physics and Astronomy, Texas A\&M
               University-Commerce, Commerce, TX 75429, USA}
\author{C.~J. Horowitz}\email{horowit@indiana.edu}
\affiliation{Center for Exploration of Energy and Matter and
                  Department of Physics, Indiana University,
                  Bloomington, IN 47405, USA}
\date{\today}
\begin{abstract}
\begin{description}

\item[Background] Pulsar glitches---the sudden spin-up in the
rotational frequency of a neutron star---suggest the existence of an
angular-momentum reservoir confined to the inner crust of the neutron
star. Large and regular glitches observed in the Vela pulsar have
originally constrained the fraction of the stellar moment of inertia
that must reside in the solid crust to about 1.4\%. However, crustal
entrainment---which until very recently has been ignored---suggests
that in order to account for the Vela glitches, the fraction of the
moment of inertia residing in the crust must increase to about
7\%. This indicates that the required angular momentum reservoir may
exceed that which is available in the crust.

\item[Purpose] We explore the possibility that uncertainties in the
equation of state provide enough flexibility for the construction of
models that predict a large crustal thickness and consequently a large
crustal moment of inertia.

\item[Methods] Moments of inertia---both total and crustal---are
computed in the slow-rotation approximation using a relativistic mean
field formalism to generate the equation of state of neutron-star
matter.

\item[Results] We compute the fractional moment of inertia of neutron
stars of various masses using a representative set of relativistic
mean-field models.  Given that analytic results suggest that the
crustal moment of inertia is sensitive to the transition pressure at
the crust-core interface, we tune the parameters of the model to
maximize the transition pressure, while still providing an excellent
description of nuclear observables. In this manner we are able to
obtain fractional moments of inertia as large as 7\% for neutron stars
with masses below 1.6 solar masses.

\item[Conclusions] We find that uncertainties in the equation of state
of neutron-rich matter are large enough to accommodate theoretical
models that predict large crustal moments of inertia. In particular, we
 find that if the neutron-skin thickness of ${}^{208}$Pb falls within the
(0.20-0.26)\,fm range, large enough transition pressures can be
generated to explain the large Vela glitches---without invoking an
additional angular-momentum reservoir beyond that confined to the
solid crust. Our results suggest that \emph{the crust may be enough}.

\end{description}
\end{abstract}

\smallskip
\maketitle

\section{Introduction}
\label{Introduction}

Rotation-powered pulsars tend to spin down slowly and steadily due
to the emission of magnetic dipole radiation, making pulsars one of
nature's most accurate clocks. For example, the Crab pulsar---with a
rotational period of $P\! \approx \!33\,$ms---spins down at a rate of 
$\dot{P}/P\! \approx\!1.3\!\times\!10^{-11}\,{\rm s}^{-1}$ or about 
$13\,\mu{\rm s}$ per year. However, in spite of this remarkable 
regularity, young pulsars often display ``glitches'' which represent 
a sudden and abrupt spin-up in their rotational frequency. Through 
high-precision pulsar timing, an extensive glitch catalogue is now 
available which, at the time of this writing, has recorded a total of 
439 glitches from 150 different pulsars\,\cite{Espinoza:2011pq,JBPGC}. 
Moreover, pulsar timing has revealed that glitches are recurrent---with 
some of the more active \emph{glitchers} being the Vela with 17, the 
Crab with 25, and PSR B1737-30 with 33\,\cite{JBPGC}.

The glitch mechanism is intimately related to the formation of
superfluid vortices in the inner crust of the rotating neutron
star\,\citep[see Refs.][and references contained
therein]{Anderson:1975zze,Pines:1985kz}. Given that many pulsars
rotate extremely fast, the areal density of superfluid vortices may be
extremely high. For example, in the case of the Crab pulsar the vortex
density reaches $n_{\rm v}\!=\!4m_{n}/\hbar
P\!\sim\!2\times\!10^{5}\,{\rm cm}^{-2}$, where $m_{n}$ is the neutron
mass. The vortex density is so high that, although the bulk superfluid
does not rotate, the superfluid as a whole appears to be rotating
collectively as a ``rigid'' body\,\cite{Pines:1985kz}. However, as the
pulsar slows down by the emission of magnetic dipole radiation, the
initial distribution of vortices---which are believed to be pinned to
a crystal lattice of neutron-rich nuclei---falls out of
equilibrium. This induces a differential rotation between the slower
neutron star and the faster superfluid vortices. When the differential
lag is too large, then suddenly and abruptly some fraction of the
vortices unpin, migrate outwards, and transfer their angular momentum
to the solid crust---and to any stellar component strongly coupled to
it (such as the liquid core). This sudden (typically in less than a
day) transfer of angular momentum is detected as a pulsar spin-up or
a \emph{glitch}. As a result of the glitch, the density of vortices
diminishes from its pre-glitch value leaving the superfluid in close
equilibrium with the solid crust. As the star continues to slow down
over a period of days to years, stresses between the crust and the
superfluid start to build up again until eventually more vortices
unpin, transfer their angular momentum to the solid crust, and
ultimately generate another glitch. The recurrence of glitches and the
ensuing recovery is therefore a manifestation of a ``sluggish" solid
crust falling periodically out of equilibrium with the larger
distribution of superfluid vortices.

Glitches from the Vela pulsar (B0833-45) were first observed in 1969
and have been continuously recorded for more than 40 years. So far a
total of 17 glitches have been documented with individual spin-up
rates of the order $\Delta\Omega/\Omega\!=\!10^{-6}$; note that
although more glitches have been recorded for the Crab pulsar, the
typical spin-up rate is three orders of magnitude
smaller\,\cite{Espinoza:2011pq,JBPGC}. The high recurrence rate and
large magnitude of the Vela glitches have been used to constrain the
underlying equation of state (EOS) of neutron-star matter in terms
of three measured quantities: its spin frequency 
$\Omega\!=\!70.338\,060\,018\,{\rm s}^{-1}$, its average spin-down
rate  $\dot{\Omega}\!=\!-9.8432\!\times\!10^{-11}$, and its 
\emph{glitch activity parameter} $A_{g}$---defined in terms of the 
cumulative spin-up rate 
$\sum_{n}\Delta\Omega_{n}/\Omega\!=\! 2.9760\!\times\!10^{-5}$  
as follows\,\cite{Link:1999ca,Chamel:2012ae}:
\begin{equation}
  A_{g} = \frac{1}{t}\sum_{n=1}^{N} \frac{\Delta\Omega_{n}}{\Omega}
  \approx 2.277 \!\times\!10^{-14}\,{\rm s}^{-1} \;,
 \label{VelaLimit1}
\end{equation}
where $t\!=\!41.421$ years is the total time elapsed between the first
and the last ($N\!=\!17$) glitches.  Thus, in the standard model of pulsar
glitches, the ratio of the moment of inertia of the superfluid
component driving the glitches $I_{s}$ to the moment of inertia of the
solid crust $I_{c}$---\emph{plus any portion of the star strongly
coupled to it}---must satisfy the following
inequality\,\cite{Link:1999ca}:
\begin{equation}
  \frac{I_{s}}{I_{c}} \ge A_{g}\frac{\Omega}{|\dot{\Omega}|} \approx 0.016.
  \label{VelaLimit2}
 \end{equation}
That is, long term timing observations of the Vela pulsar suggest that in 
order to account for its glitch activity, at least 1.6\% of the stellar moment 
of inertia must reside in the superfluid 
reservoir\,\cite{Link:1999ca,Chamel:2012ae,Hooker:2013fda}. Moreover, 
by adopting two plausible assumptions, one may rewrite the above limit
in such a way as to provide a more meaningful constraint on the EOS.
First, in the two-component model of \citet{Baym:1969}, it is
assumed that the component of the moment of inertia that rotates at
the observed stellar frequency consists of the solid outer crust
\emph{plus the liquid interior}; then,
$I_{c}\!=\!I\!-\!I_{s}\!\approx\!I$, where $I$ is the \emph{total}
stellar moment of inertia. Second, given that the inner crust is
both thicker and denser than the outer crust, the crustal moment of
inertia may be approximated as $I_{\rm crust}\!=\!I_{\rm
outer}\!+\!I_{s}\!\approx\!I_{s}$. Thus, one may express the above
constraint as follows:
\begin{equation}
   \frac{I_{s}}{I_{c}} \simeq \frac{I_{\rm crust}}{I}\gtrsim\! 0.016 \,.
 \label{VelaLimit3}
\end{equation}
This expression is particularly convenient as both $I_{\rm crust}$
and $I$ may be readily evaluated in the \emph{slow-rotation
approximation} (see Sec.\ref{Formalism}).  Indeed, most of the
formalism relies on the solution of the Tolman-Oppenheimer-Volkoff
equation whose only required input is the EOS of
neutron-star matter. Further, whereas the total moment of inertia
depends sensitively on the poorly constrained high-density component
of the EOS, the crustal component $I_{\rm crust}$ is sensitive to
physics that may be probed in the laboratory. In fact, analytic
expressions for $I_{\rm crust}$ already exist that are both
illuminating and highly accurate\,\citep[see Refs.][and references
contained therein]{Link:1999ca,Lattimer:2006xb,Fattoyev:2010tb}. In
particular, the crustal moment of inertia is highly sensitive to
physical observables---particularly the transition pressure---at the
crust-core interface.

Recently, however, the standard glitch mechanism has been called into
question\,\cite{Andersson:2012iu,Chamel:2012ae}. It has been argued
that \emph{crustal entrainment}---the non-dissipative elastic
scattering of unbound neutrons by the crystal lattice---effectively
reduces the angular-momentum reservoir. That is, entrainment effects
reduce the density of superfluid neutrons that could eventually become
the source of the superfluid vortices. The impact from crustal
entrainment may be encoded in an effective neutron mass
$m_{n}^{\star}\!\equiv\!m_{n}n_{n}^{\rm f}/n_{n}^{\rm c}$, that
reflects the ratio of unbound neutrons $n_{n}^{\rm f}$ ({\sl i.e.,}
neutrons not bound to the nuclear lattice) to those $n_{n}^{\rm c}$
neutrons that are effectively free ({\sl i.e.,} not
entrained)\,\cite{Chamel:2011aa,Chamel:2012ae}. As a result, crustal
entrainment modifies the constraint given in Eq.\,(\ref{VelaLimit3})
to
\begin{equation}
   \frac{I_{\rm crust}}{I}\gtrsim\!0.016\left(
   \frac{\langle m_{n}^{\star}\rangle}{m_{n}}\right)\simeq 0.07\;,
 \label{VelaLimit4}
\end{equation}
where in the last expression we have adopted $\langle
m_{n}^{\star}\rangle/m_{n}\!\simeq\!4.3$ as suggested in
Ref.\,\cite{Andersson:2012iu}. Note that the amount of
crustal entrainment may be uncertain, so more sophisticated
calculations could yield a different constraint for the fractional 
moment of inertia. Here we assume, as in 
Ref.\,\cite{Andersson:2012iu}, that its value lies between 
the two limits given in Eqs.\,(\ref{VelaLimit3}) and \,(\ref{VelaLimit4}); 
that is, $0.016\!\lesssim I_{\rm crust}/I\!\lesssim\,0.07$.
Given the significant impact that crustal entrainment may have
in constraining $I_{\rm crust}/I$, it has been suggested that 
\emph{the crust is not enough}, so that the core superfluid must 
also participate in glitches\,\cite{Andersson:2012iu}. However, before 
completely dismissing the standard glitch mechanism in favor of a 
more exotic explanation, we explore the conservative alternative 
that uncertainties in the EOS (which are large) could still allow the 
inner crust to  account---by itself---for a large enough fraction of 
the moment of inertia to explain the large Vela glitches.

Uncertainties in the EOS are known to significantly affect the
transition between the liquid interior and the solid crust.
Particularly relevant to the present discussion is the poorly-known
density dependence of the symmetry energy $S(\rho)$.  The symmetry
energy represents the energy required to convert symmetric nuclear
matter into pure neutron matter at a fixed baryon density $\rho$.
Although the symmetry energy at saturation density is fairly well
determined, the value of its slope $L$ remains highly uncertain. And
it is precisely the slope of the symmetry energy that controls the
properties of neutron-rich matter at the crust-core transition
region\,\cite{Horowitz:2000xj}. Indeed, the transition density and
proton fraction at the crust-core interface are both linearly
anti-correlated to $L$. However, in stark contrast, the transition
pressure does not vary monotonically with
$L$\,\cite{Ducoin:2010as,Fattoyev:2010tb}. Note that
such lack of correlation between the transition pressure and
$L$---which appears to contradict some recent studies\,\cite{Worley:2008cb,
Xu:2008vz,Xu:2009vi,Moustakidis:2010zx,Paar:2014qda}---emerges
only as one systematically explores a wide range of models. Given
that the crustal moment of inertia grows with an increasing
transition pressure, we want to explore the possibility of
generating realistic EOS that predict large transition pressures. As
we shall demonstrate, we find a class of relativistic mean field
models with moderate values of $L$ that generate fractional moments
of inertia as large as $I_{\rm crust}/I\!\approx\!0.09$ for a
canonical $1.4\,M_{\odot}$ neutron star. By using
neutron star observations, Ref.\,\cite{Steiner:2014pda} has recently
and independently reached a similar conclusion. Moreover, we argue
that these predictions may be directly tested in the laboratory.
Although $L$ can not be directly measured in the laboratory, it is
known to be strongly correlated to the neutron-skin thickness of
${}^{208}$Pb\,\cite{Brown:2000,
Furnstahl:2001un,Centelles:2008vu,RocaMaza:2011pm}---a fundamental
nuclear-structure observable that has been
measured\,\cite{Abrahamyan:2012gp,Horowitz:2012tj}---and will be
measured with increasing accuracy---at the Jefferson
Laboratory\,\cite{PREXII:2012}. In summary, it is the main goal of
the present work to explore whether present uncertainties in the EOS
are large enough to accommodate realistic models that could account
for large pulsar glitches---even in the presence of crustal
entrainment.

We have organized the paper as follows. In Sec.\,\ref{Formalism} we
review the essential details required to compute the stellar moment of
inertia and the class of equations of state that will be used. In
particular, special attention is paid to the crustal component of the
moment of inertia and its sensitivity to the transition pressure at
the crust-core interface. In Sec.\,\ref{Results} we provide predictions
for the fraction of the moment of inertia residing in the crust. We
show that although the transition pressure is sensitive to the slope
of the symmetry energy, its behavior is not monotonic. This suggests
a range  of values for $L$---that are neither too small nor too
large---that provide the thickest crust and consequently the
largest fractional moment of inertia. Finally, we offer our conclusions
in Sec.\,\ref{Conclusions}.

\section{Formalism}
\label{Formalism}

In this section we provide a brief review of the formalism required to
compute the moment of inertia of a neutron star. First, we outline the
procedure required to compute the stellar moment of inertia in the
\emph{slow-rotation} approximation. The enormous advantage of this
approach is that most of what is required is a solution of the
Tolman-Oppenheimer-Volkoff (TOV) equations in the limit of
\emph{spherical} symmetry. Given that the neutron-star matter equation
of state is the sole ingredient required to solve the TOV equations,
we devote a second subsection to review the main physical assumptions
underlying our models.

\subsection{Moment of inertia of a neutron star}
\label{Sec:MomIn}
In the slow-rotation approximation pioneered by Hartle and
Thorne\,\cite{Hartle:1967he,Hartle:1968si} the moment of inertia of a
uniformly rotating, axially symmetric neutron star is given by the
following expression:
\begin{equation}
  I \equiv \frac{J}{\Omega} = \frac{8\pi}{3}
 \int_{0}^{R} r^{4} e^{-\nu(r)}\frac{\bar{\omega}(r)}{\Omega}
 \frac{\Big({\mathcal E}(r)+P(r)\Big)}{\sqrt{1-2GM(r)/r}} dr \;,
 \label{MomInertia}
\end{equation}
where $J$ and $\Omega$ are the angular momentum and rotational
frequency of the neutron star.  Further, $M(r)$, ${\mathcal E}(r)$,
and $P(r)$ are the stellar mass, energy density, and pressure
profiles, respectively. Finally, $\nu(r)$ and $\omega(r)$ (with
$\bar{\omega}(r)\!\equiv\!\Omega\!-\!\omega(r)$) are
radially-dependent metric functions. In the absence of general
relativistic effects and in the simplified situation of a neutron star
of uniform density, the stellar moment of inertia reduces to the
elementary result of $I\!=\!2MR^{2}/5$. However, for realistic
situations the stellar moment of inertia differs considerably from
this elementary result.  Note that in order for the slow-rotation
approximation to be valid, the stellar frequency $\Omega$ must be
far smaller than the mass-shedding Kepler frequency $\Omega_{\rm K}$.
That is,
\begin{equation}
  \Omega\!\ll\!\Omega_{\rm K}\!=\!\sqrt{\frac{GM}{R^{3}}}\!\approx\!
  (1.15\!\times\!10^{4}\,{\rm s}^{-1}) \sqrt{\frac{M/M_{\odot}}{(R/R_{10})^{3}}}\;,
 \label{OmKepler}
\end{equation}
where in the last expression the stellar mass is expressed in solar
masses and the radius in units of $R_{10}\!\equiv\!10\,{\rm km}$.
Equivalently, the slow-rotation approximation is valid if the period
of the neutron star is significantly longer than about half a millisecond.
Both the Crab and Vela pulsars---with periods of 33 and
89\,milliseconds, respectively---fall safely within this bound.

The enormous virtue of the slow-rotation approximation is that every
quantity that appears in Eq.~(\ref{MomInertia}) may be computed in the
limit of spherical symmetry. As such, with the exception of the
\emph{Lense-Thirring frequency} $\omega(r)$, they may all be
obtained from a solution of the Tolman-Oppenheimer-Volkoff
(TOV) equations---properly supplemented by an equation of state
$P\!=\!P({\mathcal E})$. However, one should note that even under
this simplified situation the formalism remains complex. Indeed,
the Lense-Thirring precession is a qualitative new effect with no
counterpart in Newtonian gravity that is caused by the dragging
of inertial frames around a rotating compact star. Nevertheless,
once the TOV equations are solved so that mass, energy
density, and pressure profiles are obtained, the remaining quantities
appearing in Eq.~(\ref{MomInertia}) may be computed by either performing
a quadrature (in the case of $\nu$) or by solving a suitable differential
equation (in the case of $\bar{\omega}$)\,\cite{Fattoyev:2010tb}. The
formalism for the moment of inertia of an axisymmetric star in hydrostatic
equilibrium is derived in far greater detail in
Refs.\,\cite{Hartle:1973zza,Weber:1999}, although a more pedagogical
discussion may be found in the text by Glendenning\,\cite{Glendenning:2000}.

Whereas the evaluation of the stellar moment of inertia relies on the
numerical computation of various quantities, accurate and illuminating
analytic expressions for the crustal moment of inertia already
exist. The crustal component of the moment of inertia is defined in
terms of the integral given in Eq.~(\ref{MomInertia}) but now with the
integration range limited from the transition (or core) radius $R_{t}$
to the stellar radius $R$. That is,
\begin{equation}
  I_{\rm crust} = \frac{8\pi}{3}
 \int_{R_{t}}^{R} r^{4} e^{-\nu(r)}\widetilde{\omega}(r)
 \frac{\Big({\mathcal E}(r)+P(r)\Big)}{\sqrt{1-2GM(r)/r}} dr \;.
 \label{Icr0}
\end{equation}
However, given that the crust is thin and the density within it is
low, a closed-form approximation for $I_{\rm crust}$ has been
obtained\,\cite{Lorenz:1992zz,Ravenhall:1994,Link:1999ca,Lattimer:2000nx,
Lattimer:2006xb,Fattoyev:2010tb}. That is,
\begin{equation}
  I_{\rm crust} \approx \frac{16\pi}{3}\frac{R_{t}^{6}P_{t}}{R_{s}}
 \left[1-\frac{0.21}{(R/R_{s}-1)}\right]
 \left[1+\frac{48}{5}(R_{t}/R_{s}-1)(P_{t}/{\mathcal E}_{t}) +
   \ldots\right] \;.
 \label{Icr1}
\end{equation}
where $R_{s}\!=\!2GM$ is the Schwarzschild radius of the star, and
$P_{t}$ and ${\mathcal E}_{t}$ are the pressure and energy density at
the crust-core interface. The {\sl ellipsis} in the above equation indicates
that the derivation was carried out to first order in the small (typically
$\lesssim\!1\%$) quantity $P_{t}/{\mathcal E}_{t}$. Note that the TOV
equation in the crustal region may be solved exactly using a polytropic
EOS\,\cite{Fattoyev:2010tb}.
Although in reporting results for the crustal moment of inertia the exact
expression given in Eq.\,(\ref{Icr0}) will always be used, the above analytic
expression provides valuable insights. First, although approximate, the
expression has been shown to be highly accurate\,\cite{Fattoyev:2010tb}.
Second, the two terms that appear between brackets in Eq.\,(\ref{Icr1}) provide
small ($\sim$10\%) and canceling contributions to the leading term. Thus,
the crustal moment of inertia of a neutron star of mass $M$ is
dominated by $R_{t}\!\equiv\!R\!-\!R_{\rm crust}$ and $P_{t}$---quantities
that are highly sensitive to the poorly-constrained density dependence of
the symmetry energy. Note that the low-density environment near the
crust-core interface may be simulated in the collisions of heavy ions.
Indeed, novel and unique experiments involving heavy-ion collisions
with very neutron-rich nuclei at both present and future radioactive beam
facilities will provide significant constraints on the EOS of
neutron-rich matter\,\cite{Horowitz:2014bja}.

\subsection{Equation of state for neutron-star matter}
\label{Sec:EOS}
Neutron stars are rich dynamical systems with core densities and
neutron-proton asymmetries that far exceed those found in nuclei
under normal conditions. As such, the EOS required to understand
the structure and dynamics of neutron stars is highly uncertain and
involves uncontrolled extrapolations. To mitigate this problem we
rely on several nuclear-structure models that have been previously
constrained by a variety of laboratory observables. These models,
that fall under the rubric of \emph{relativistic mean-field models},
provide a Lorentz covariant and causal framework that becomes
essential as one extrapolates to the high densities encountered in
the stellar core\,\cite{Walecka:1974qa,Serot:1984ey,Serot:1997xg}.

The EOS for the uniform liquid core is based on an
interacting Lagrangian density with its parameters accurately
calibrated to a variety of nuclear properties. Besides nucleons, the
uniform liquid core contains leptons---both electrons and
muons---that enforce charge neutrality and chemical equilibrium
and are treated as relativistic free Fermi gases. However, the
model contains no exotic degrees of freedom. The interacting
Lagrangian density includes a nucleon field ($\psi$) interacting via
the exchange of two isoscalar mesons (a scalar $\phi$ and a vector
$V^{\mu}$) and one isovector meson
($b^{\mu}$)\,\cite{Serot:1984ey,Serot:1997xg}. That is, 
\begin{equation}
{\mathscr L}_{\rm int} =
\bar\psi \left[g_{\rm s}\phi   \!-\!
         \left(g_{\rm v}V_\mu  \!+\!
    \frac{g_{\rho}}{2}\mbox{\boldmath$\tau$}\cdot{\bf b}_{\mu}
          \right)\gamma^{\mu} \right]\psi -
          U(\phi,V^{\mu},{\bf b^{\mu}}) \;,
\label{Lagrangian}
\end{equation}
where $g_{\rm s}$, $g_{\rm v}$, and $g_{\rho}$ represent the Yukawa
couplings between the nucleon and the corresponding meson fields.
Although essential for the description of the properties of finite nuclei,
note that the photon field does not contribute to the EOS at the
mean-field level. In order to improve the phenomenological standing
of the model, the Lagrangian density must be supplemented by nonlinear
meson interactions that introduce an effective density dependence.
These non-linear terms are given by
\begin{equation}
  U(\phi,V^{\mu},{\bf b}^{\mu}) =
   \frac{\kappa}{3!} (g_{\rm s}\phi)^3 \!+\!
    \frac{\lambda}{4!}(g_{\rm s}\phi)^4  \!-\!
   \frac{\zeta}{4!}
    \Big(g_{\rm v}^2 V_{\mu}V^\mu\Big)^2 \!-\!
    \Lambda_{\rm v}
    \Big(g_{\rho}^{2}\,{\bf b}_{\mu}\cdot{\bf b}^{\mu}\Big)
    \Big(g_{\rm v}^2V_{\nu}V^\nu\Big) \;.
\label{USelf}
\end{equation}
Historically, the first two terms ($\kappa$ and $\lambda$) were
introduced by Boguta and Bodmer to soften the compressibility
coefficient of symmetric nuclear matter\,\cite{Boguta:1977xi}. The
last two terms ($\zeta$ and $\Lambda_{\rm v}$) on the other hand, play
a critical role in the behavior of the EOS at high densities. In
particular, the isoscalar parameter $\zeta$ serves to soften the
equation of state of symmetric nuclear matter at high
densities\,\cite{Mueller:1996pm} and, at present, can only be
meaningfully constrained by the maximum mass of a neutron
star\,\cite{Demorest:2010bx,Antoniadis:2013pzd}. Finally, the
\emph{mixed isoscalar-isovector} parameter $\Lambda_{\rm v}$ was
introduced to modify the poorly-constrained density dependence of the
symmetry energy\,\cite{Horowitz:2000xj,Horowitz:2001ya}. Tuning this
parameter provides a simple, efficient, and reliable method of
softening the symmetry energy. Further details on the calibration
procedure may be found in Refs.\,\cite{Horowitz:2000xj,
Todd:2003xs,Fattoyev:2013yaa} and references contained therein.

As one moves away from the stellar core the density diminishes until
the uniform system becomes unstable against cluster formation. That
is, at densities of about one half of nuclear-matter saturation
density, it becomes energetically favorable for the uniform system to
cluster into neutron-rich fragments that are embedded in a dilute
neutron vapor.  Such instability delineates the transition between the
uniform liquid core and the non-uniform solid crust. The stellar crust
itself is divided into an inner and an outer region. In the outer
crust the system is organized into a Coulomb lattice of neutron-rich
nuclei embedded in a degenerate electron
gas\,\cite{Baym:1971pw,RocaMaza:2008ja}. The composition in this
region is solely determined by the masses of neutron-rich nuclei in
the $26\!\le\!Z\lesssim\!50$ region and the pressure support is
provided by the degenerate electrons. For this region we adopt the
standard equation of state of Baym, Pethick, and Sutherland
(BPS)\,\cite{Baym:1971pw}. In contrast, the EOS for the inner crust is
highly uncertain and must be inferred from theoretical calculations.
In addition to a Coulomb lattice of progressively more exotic
neutron-rich nuclei embedded in an electron gas, the inner crust now
includes a dilute vapor of superfluid neutrons. Moreover, at the
bottom layers of the inner crust, complex and exotic structures with
almost equal energy---``nuclear pasta"---have been predicted to
emerge\,\cite{Ravenhall:1983uh,Hashimoto:1984,Lorenz:1992zz}.
Whereas significant progress has been made in simulating this exotic
region\,\cite{Watanabe:2003xu,Watanabe:2004tr,Horowitz:2004yf,
Horowitz:2004pv,Horowitz:2005zb,Schneider:2013dwa}, a detailed
equation of state is still missing. Hence, we resort to a polytropic
EOS to interpolate between the solid outer crust and the uniform
liquid interior\,\cite{Link:1999ca,Carriere:2002bx}.

In summary, the equation of state adopted in this work is comprised of
three parts: (i) a BPS component up to a neutron-drip density of
$\rho_{\rm drip}\!\approx\!4\!\times\!10^{11}\,{\rm g/cm^{3}}$ to
describe the outer crust; (ii) a polytropic component for the inner
crust that interpolates between $\rho_{\rm drip}$ and the crust-core
transition density $\rho_{t}$; and (iii) a RMF component to describe
the uniform liquid core. Note that $\rho_{t}$ is a model-dependent
quantity highly sensitive to the density dependence of the symmetry
energy. Consistency demands that the same RMF model used to generate
the EOS will also be used to predict the crust-core transition
density. In particular, the instability of the uniform ground state
against cluster formation will be determined through a relativistic
random phase approximation (RPA) as detailed in
Refs.\,\cite{Horowitz:2000xj, Carriere:2002bx}.

\section{Results}
\label{Results} 
\begin{figure*}[htbp]
 \begin{center}
  \includegraphics[width=0.47\textwidth]{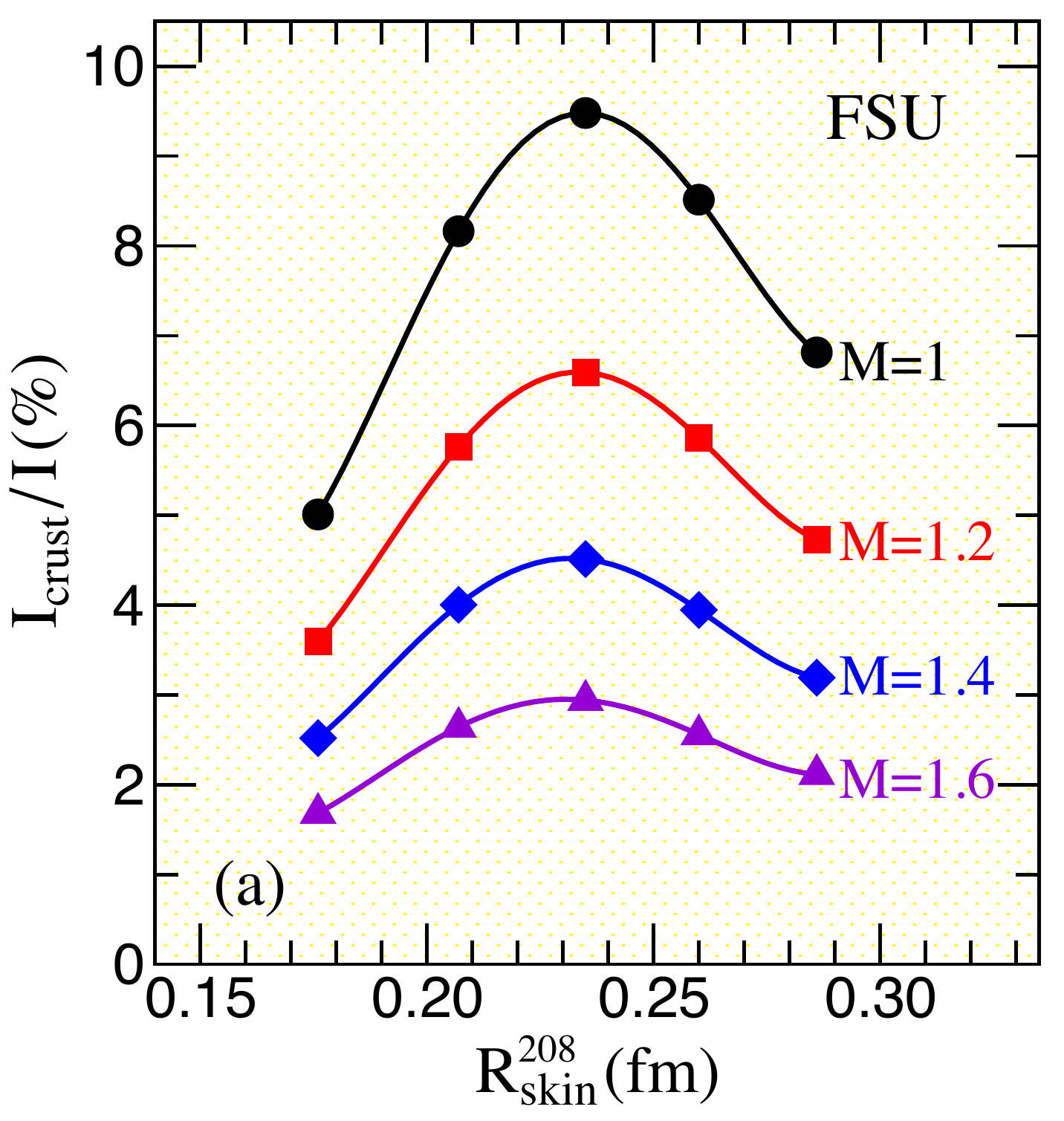}
        \hspace{0.1cm}
  \includegraphics[width=0.47\textwidth]{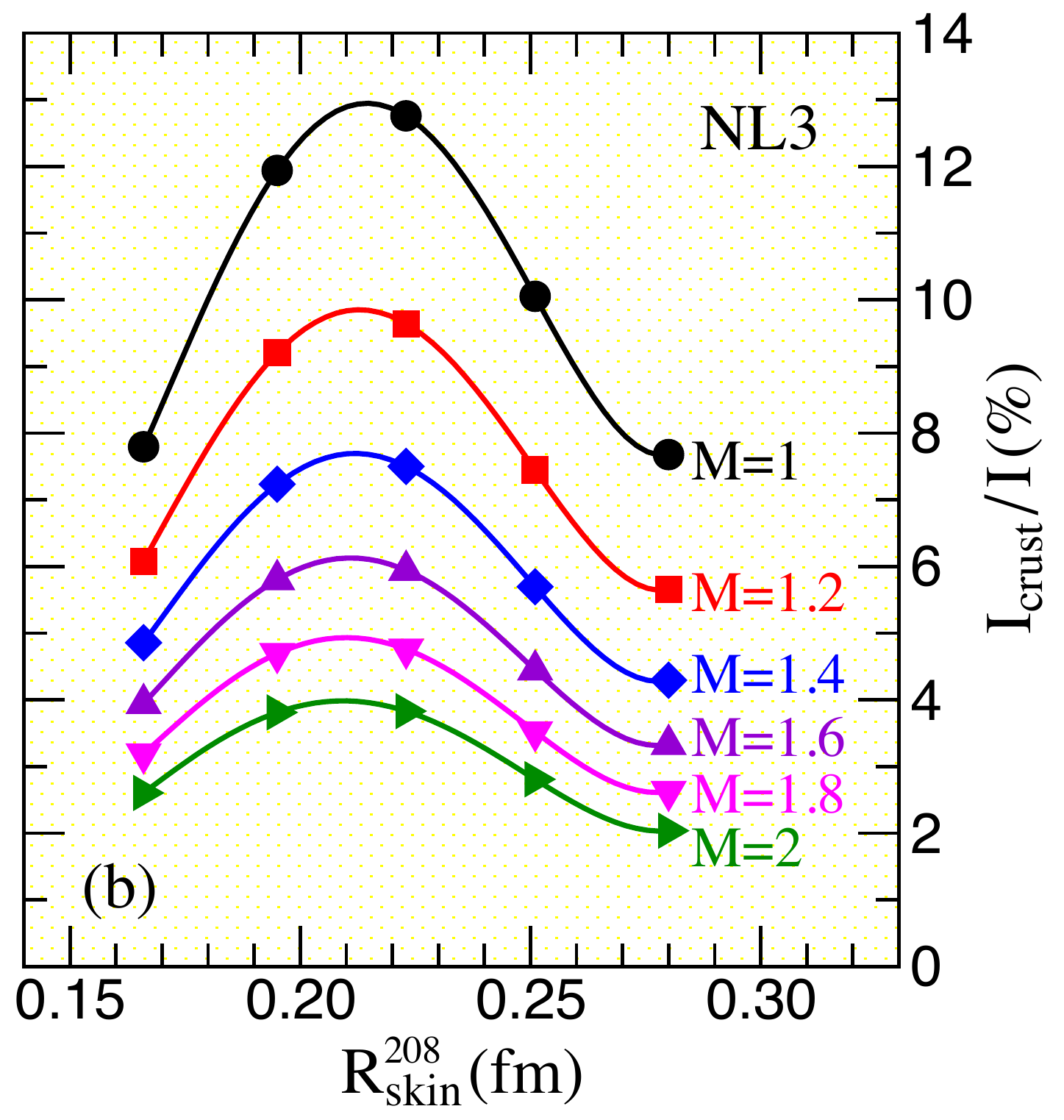}
  \caption{(Color online)
  Fraction of the crustal moment of inertia as a function of the neutron-skin
  thickness of ${}^{208}$Pb for a variety of neutron-star masses (in units of
  the solar mass) as predicted by two ``families'' of mean-field models. (a)
  Predictions made with the relatively \emph{soft} FSU\,\cite{Todd-Rutel:2005fa}
  parametrization and (b) with the relatively \emph{stiff} NL3
  interaction\,\cite{Lalazissis:1996rd,Lalazissis:1999}.}
 \label{Fig1}
 \end{center}
\end{figure*}

In this section we display results for the fraction
of the stellar moment of inertia residing in the non-uniform crust.
By using a variety of realistic equations of state, we test the
basic premise of our work, namely, whether \emph{the crust may be
enough} to explain large pulsar glitches.  We start by
displaying in Fig.\,\ref{Fig1} the fraction of the crustal moment of
inertia for a variety of neutron-star masses for two ``families" of
relativistic mean-field models: FSUGold\,\cite{Todd-Rutel:2005fa}
(or ``FSU" for short) and
NL3\,\cite{Lalazissis:1996rd,Lalazissis:1999}. Note that each member
of the family is characterized by a unique value of the neutron-skin
thickness of ${}^{208}$Pb. The original two models---with values of
$R_{\rm skin}^{208}\!=\!0.21\,{\rm fm}$ for FSU and $R_{\rm
skin}^{208}\!=\!0.28\,{\rm fm}$ for NL3---were accurately calibrated
to several nuclear properties and have been fairly successful in
describing a variety of nuclear phenomena. Yet, from the perspective
of the EOS, a critical difference between these two
models is that NL3 predicts a significantly stiffer EOS than FSU. In
particular, NL3 predicts a maximum neutron star mass almost one
solar mass heavier than FSU. Moreover, with a symmetry energy also
significantly stiffer, NL3 predicts considerably larger stellar
radii. However, the density dependence of the symmetry energy may be
efficiently modified by tuning the isoscalar-isovector coupling
$\Lambda_{\rm v}$. Indeed, it is precisely through the tuning of
$\Lambda_{\rm v}$ that we have generated the two families of RMF
models displayed in Fig.\,\ref{Fig1}. Note that in doing so, all
isoscalar properties---such as the EOS of symmetric nuclear
matter---are left intact\,\cite{Fattoyev:2010tb}. Also note that we
have quantified the sensitivity of the crustal fraction of the
moment of inertia to the density dependence of the symmetry energy
in terms of $R_{\rm skin}^{208}$.  This provides an attractive
connection between a laboratory measurement and an astrophysical
observation.

The lack of a linear correlation between the neutron skin thickness
$R_{\rm skin}^{208}$ and the fraction of the crustal moment of inertia
is clearly evident in Fig.\,\ref{Fig1}. Whereas earlier studies have
found a strong linear correlation between $R_{\rm skin}^{208}$ and
various properties of the EOS in the crust-core transition
region\,\cite{Horowitz:2000xj,Fattoyev:2010tb}---such as the baryon
density, energy density, and proton fraction---no such correlation was
observed in the case of the transition
pressure\,\cite{Ducoin:2010as,Fattoyev:2010tb}. Given the sensitivity
of the crustal moment of inertia to the transition pressure, as
indicated in Eq.\,(\ref{Icr1}), we now use such lack of correlation to
identify the particular member of the FSU and NL3 families that yields
the largest transition pressure and, consequently, the largest
fraction of the crustal moment of inertia. As suggested by the results
displayed in Fig.\,\ref{Fig1}, this occurs for models with
``intermediate'' values for the neutron-skin thickness of
${}^{208}$Pb.

\begin{table}[h]
\begin{tabular}{|l||c|c|c|c|c|c|c|c|c|l|}
 \hline
 Model & $m_{\rm s}$  & $m_{\rm v}$  & $m_{\rho}$
       & $g_{\rm s}^2$ & $g_{\rm v}^2$ & $g_{\rho}^2$
       & $\kappa$ & $\lambda$ & $\zeta$ & $\hfil\Lambda_{\rm v}$\\
 \hline
 \hline
 IU-FSU       & 491.500 & 782.500 & 763.000 &  99.4266 & 169.8349 & 184.6877
                   & 3.3808  & $+2.9600\!\times\!10^{-4}$ & 0.0300 & $4.6000\!\times\!10^{-2}$\\[-0.05cm]
 FSUmax    & 491.500 & 782.500 & 763.000 & 112.1996 & 204.5469 & 122.9556
                   & 1.4203  & $+2.3762\!\times\!10^{-2}$& 0.0600 & $2.4776\!\times\!10^{-2}$\\[-0.05cm]
 NL3max     & 508.194 & 782.501 & 763.000 & 104.3871 & 165.5854 &  119.3049
                   & 3.8599  & $-1.5905\!\times\!10^{-2}$& 0.0000 & {$2.6708\!\times\!10^{-2}$}\\[-0.05cm]
 TFcmax     &496.800 & 782.500 & 763.000 & 113.9565 & 198.0546 & 149.9692
                   & 2.6078  & $-1.8640\!\times\!10^{-3}$& 0.0200 & $1.5133\!\times\!10^{-2}$\\
\hline
\end{tabular}
\caption{Parameter sets for the relativistic mean-field models used in the text.
             The parameter $\kappa$ and the meson masses $m_{\rm s}$, $m_{\rm v}$,
             and $m_{\rho}$ are all given in MeV. The nucleon mass has been fixed at
             $M\!=\!939$~MeV in all models.}
\label{Table1}
\end{table}

The parameters for the four RMF models employed in this manuscript
have been listed in Table\,\ref{Table1}. We reiterate that all these
models provide a fairly accurate description of various properties of
finite nuclei throughout the nuclear chart. In particular, the IU-FSU
parametrization was derived from the original FSUGold model by
adjusting a few empirical parameters with the goal of softening the
symmetry energy (in order to generate smaller stellar radii) and
stiffening the overall EOS at higher density (to generate a larger
limiting neutron-star mass)\,\cite{Fattoyev:2010mx}. In turn, the
``FSUmax" set represents that particular member of the FSU family the
predicts the maximum value for the transition pressure $P_{t}$. The
remaining two sets, ``NL3max" and ``TFcmax", were selected in a
similar fashion from the corresponding
NL3\,\cite{Lalazissis:1996rd,Lalazissis:1999} and
TFc\,\cite{Fattoyev:2013yaa} families. That is, with the exception of
IU-FSU that was selected as an example of a model with a fairly soft
symmetry energy, the remaining three models were chosen to maximize
the value of the transition pressure.

\begin{table}[h]
\begin{tabular}{|l||c|c|c|c|c|c||c|c|c|}
 \hline
 Model & $\rho_{{}_{0}}({\rm fm}^{-3})$ & $\varepsilon_{0}$ & $K_{0}$ & $J$ & $L$ &
 $K_{\rm sym}$ & $B/A$ & $R_{\rm ch}$(fm) & $R_{\rm skin}$(fm) \\
\hline
\hline
IU-FSU    & 0.155  & $-$16.40 & 231.33 & 31.30 & 47.20
               & $+$28.53 & $-$7.89 & 5.48 & 0.16\\[-0.05cm]
FSUmax  & 0.148  & $-$16.30 & 230.01 & 33.23 & 66.00
               & $-$65.59 & $-$7.89 & 5.51 & 0.22\\[-0.05cm]
NL3max  & 0.148  & $-$16.24 & 271.54 & 32.14 & 59.00
              & $-$25.44 & $-$7.92 & 5.52 & 0.20\\[-0.05cm]
TFcmax  & 0.148  & $-$16.46 & 260.49 & 38.28 & 74.00
              & $-$67.78 & $-$7.89 & 5.51 & 0.26\\[-0.05cm]
\hline
Experiment &  &  &  &  &  &  & $-$7.87 & 5.50 & $0.33^{+0.16}_{-0.18}$ \\
\hline
\end{tabular}
\caption{Predictions for the bulk parameters characterizing the behavior of infinite
nuclear matter at saturation density $\rho_{_{0}}$. The quantities $\varepsilon_{_{0}}$
and $K_{0}$ represent the binding energy per nucleon and incompressibility coefficient
of symmetric nuclear matter, whereas $J$, $L$, and  $K_{\rm sym}$ represent the energy,
slope, and curvature of the symmetry energy. Also shown are the binding energy per
nucleon, charge radius, and neutron-skin thickness of ${}^{208}$Pb, alongside the
corresponding experimental values. All quantities are given in MeV except otherwise
indicated.}
\label{Table2}
\end{table}


Before displaying our results for various crustal properties, we discuss briefly some of 
the uncertainties in the EOS of relevance to the present work. We start by
listing in Table\,\ref{Table2} model predictions for a few bulk parameters that characterize 
the behavior of asymmetric nuclear matter in the vicinity of saturation density. That is, we
write the energy per nucleon as follows:
\begin{equation}
  \frac{E}{A}(\rho,\alpha) =
  \left(\varepsilon_{{}_{0}}+\frac{1}{2}K_{0}x^{2}+\ldots\right)+
 \left(J + Lx + \frac{1}{2}K_{\rm sym}x^{2}+\ldots\right)\alpha^{2}
 + \ldots
 \label{AsymNM}
\end{equation}
where $\alpha\!\equiv\!(N\!-\!Z)/A$ is the neutron-proton asymmetry
and $x\!\equiv\!(\rho\!-\!\rho_{{}_{0}})/3\rho_{{}_{0}}$ quantifies
the deviations of the density from its saturation
value\,\cite{Piekarewicz:2008nh,Piekarewicz:2013bea}. Moreover,
$\varepsilon_{_{0}}$ and $K_{0}$ represent the binding energy
per nucleon and incompressibility coefficient of symmetric nuclear
matter, whereas $J$, $L$, and  $K_{\rm sym}$ represent the energy,
slope, and curvature of the symmetry energy. Also shown in
Table\,\ref{Table2} are predictions for the binding energy per nucleon,
charge radius, and neutron-skin thickness of ${}^{208}$Pb---alongside
the measured experimental values. Note that the pioneering Lead Radius
Experiment (PREX) at the Jefferson Laboratory has recently provided the
first electroweak determination of the neutron radius of ${}^{208}$Pb.
Given that its charge radius has been accurately known for many years,
PREX effectively measured the neutron-skin thickness of ${}^{208}$Pb
to be\,\cite{Abrahamyan:2012gp,Horowitz:2012tj}:
\begin{equation} R_{\rm skin}^{208} =\!{0.33}^{+0.16}_{-0.18}\,{\rm fm}.
\end{equation}
The results in Table \,\ref{Table2} indicate that although the extensive
experimental database of nuclear masses and charge radii is sufficient
to constrain some of the bulk parameters of neutron-rich matter, it is
insufficient to constrain them all---particularly those associated with
the density dependence of the symmetry energy. It is precisely
this \emph{flexibility} that we have exploited in constructing theoretical
models that predict large transition pressures without compromising their
ability to describe well-measured nuclear observables.


\begin{table}
\begin{tabular}{|l||c|c|c|c||c|c|c|}
  \hline
   Model & $\rho_{t}\,({\rm fm}^{-3})$ & $Y_{t}$ & ${\mathcal E}_{t}({\rm MeV\,fm}^{-3})$
             & $P_{t}({\rm MeV\,fm}^{-3})$ & $M_{\rm crust}(M_{\odot})$ & $R_{\rm crust}$(km)
             & $I_{\rm crust}/I$(\%)\\
  \hline
  \hline
  IU-FSU    & 0.0871 & 0.0438 & 82.655 & 0.289 & 0.019 & 0.989 & 2.827\\[-0.05cm]
  FSUmax  & 0.0727 & 0.0321 & 68.810 & 0.425 & 0.027 & 1.430 & 4.407\\[-0.05cm]
  NL3max  & 0.0826 & 0.0357 & 78.290  & 0.550 & 0.052 & 1.990 & 7.619\\[-0.05cm]
  TFcmax  & 0.0794 & 0.0484 & 75.454  & 0.692 & 0.061 & 2.372 & 9.258\\
  \hline
\end{tabular}
 \caption{Predictions for the transition density, proton fraction, energy density,
 and pressure at the crust-core interface. Also shown are predictions for the
 crustal mass, radius, and fractional moment of inertia for a 1.4$M_{\odot}$
 neutron star.}
 \label{Table3}
\end{table}

\begin{figure*}[htbp]
 \begin{center}
  \includegraphics[width=0.55\textwidth]{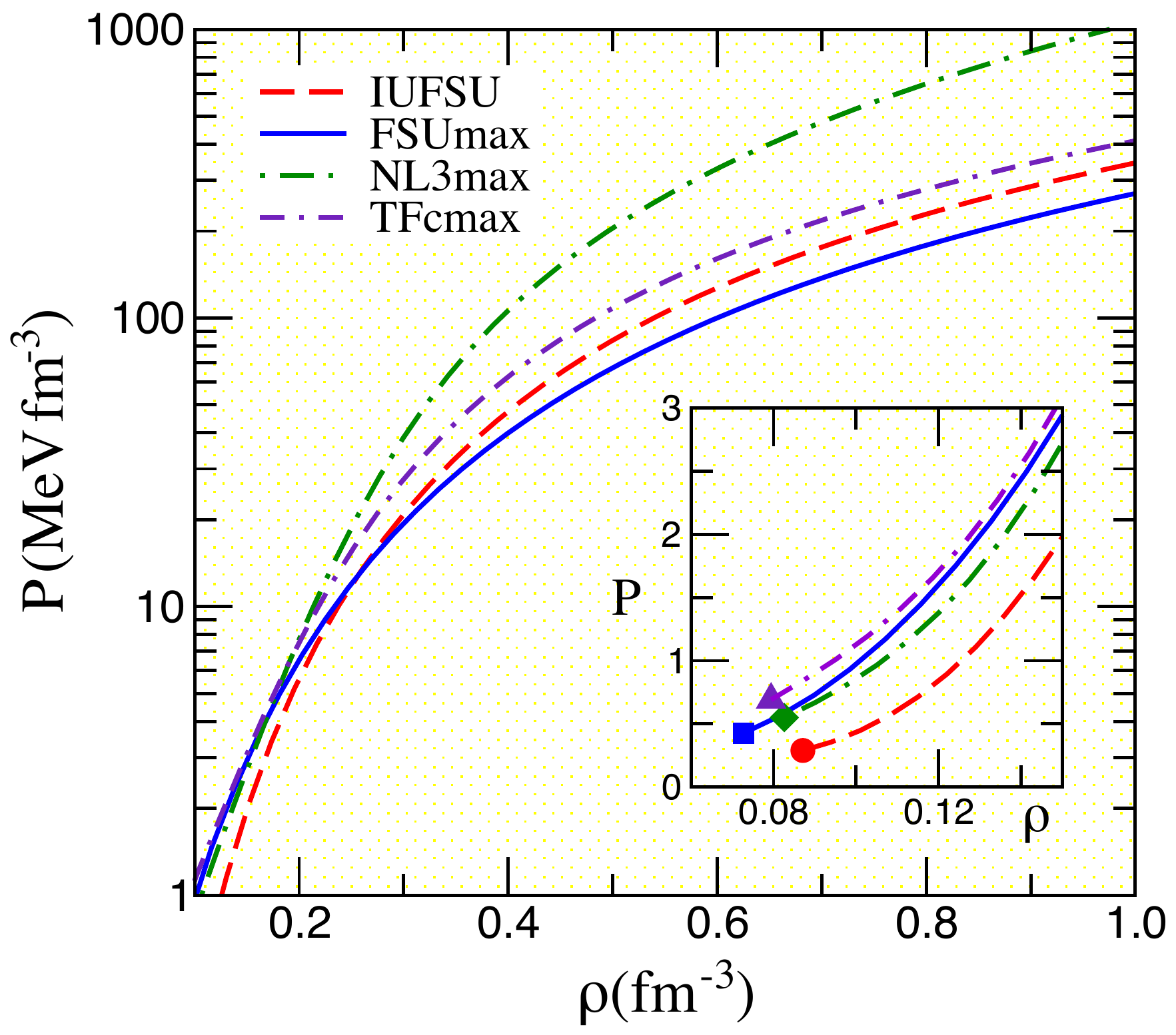}
  \caption{(Color online) Equation of state in the form of pressure-vs-baryon density
  as predicted by the four models considered in the text. The insert displays (on a
  linear scale) the EOS on a more limited range: from the crust-core transition density
  up to nuclear-matter saturation density. The symbols denote the density and pressure
  at the crust-core interface.}
 \label{Fig2}
 \end{center}
\end{figure*}

To further elucidate some of the uncertainties in the EOS that are of
relevance to the transition region between the liquid core and the
solid crust, we list in Table\,\ref{Table3} the density, proton
fraction, energy density, and pressure at the crust-core interface.
Also shown are predictions for the crustal mass, radius, and
fractional moment inertia for a ``canonical'' 1.4$M_{\odot}$ neutron
star. As expected, crustal properties are strongly correlated to the
transition pressure $P_{t}$.  Indeed, the larger the value of $P_{t}$,
the larger the fraction of the mass, radius, and moment of inertia
contained in the solid crust. To complement the tabular information we
display in Fig.\,\ref{Fig2} the EOS, {\sl i.e.,} pressure as a
function of baryon density, for uniform neutron-star matter.  Note
that as a result of the saturation of symmetric nuclear matter, the
pressure in the vicinity of saturation density ($\rho_{{}_{0}}\!\simeq
0.15\,{\rm fm}^{-3}$) is dominated by the contribution from the
symmetry energy; that is, $P\!\approx\!\rho_{{}_{0}}L/3$. Moreover,
since $L$ is strongly correlated to $R_{\rm skin}^{208}$, one can
readily infer the relative size of the neutron-skin thickness of
${}^{208}$Pb from the value of the pressure at saturation
density. Remarkably, this same pressure controls the stellar radius of
low-mass neutron stars\,\cite{Carriere:2002bx} (see
Fig.\ref{Fig5}). However, whereas stellar radii are sensitive to the
density dependence of the symmetry energy in the immediate vicinity of
nuclear-matter saturation density\,\cite{Lattimer:2006xb}, the maximum
neutron-star mass depends on the high-density component of the
EOS---which is largely controlled by the parameter $\zeta$ [see
Eq.(\ref{USelf})]. In particular, the smaller the value of $\zeta$ the
larger the limiting neutron-star mass. Thus, by tuning $\zeta$ one can
efficiently increase the maximum neutron-star mass without affecting
the EOS at normal nuclear densities\,\cite{Mueller:1996pm}.  Indeed,
for the RMF models employed here one obtains:
$M_{\rm max}/M_{\odot}\!=\!1.72, 1.94, 2.14, {\rm and\,} 2.75$, for FSUmax,
IU-FSU, TFcmax, and NL3max, respectively. Finally, the insert displays
the EOS (on a linear scale) starting from the transition point up to
saturation density.  Indicated with the various symbols are the
predictions from the different models for the crust-core transition
density and pressure. Clearly, the behavior of $P_{t}$ is not
monotonic; whereas TFcmax predicts the largest transition pressure,
its transition density is neither the smallest nor the largest.

\begin{figure*}[htbp]
 \begin{center}
  \includegraphics[width=0.47\textwidth]{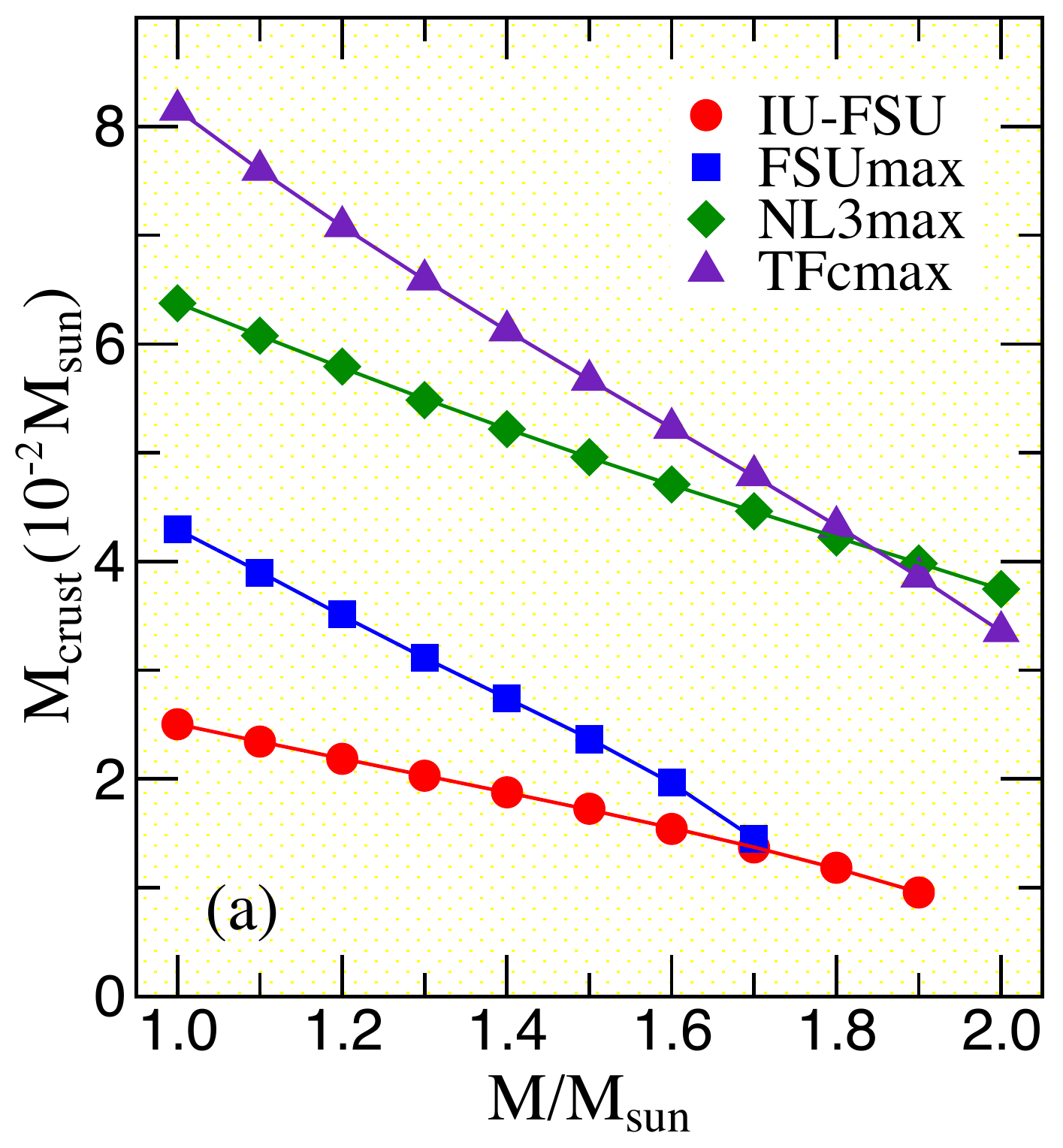}
        \hspace{0.1cm}
  \includegraphics[width=0.47\textwidth]{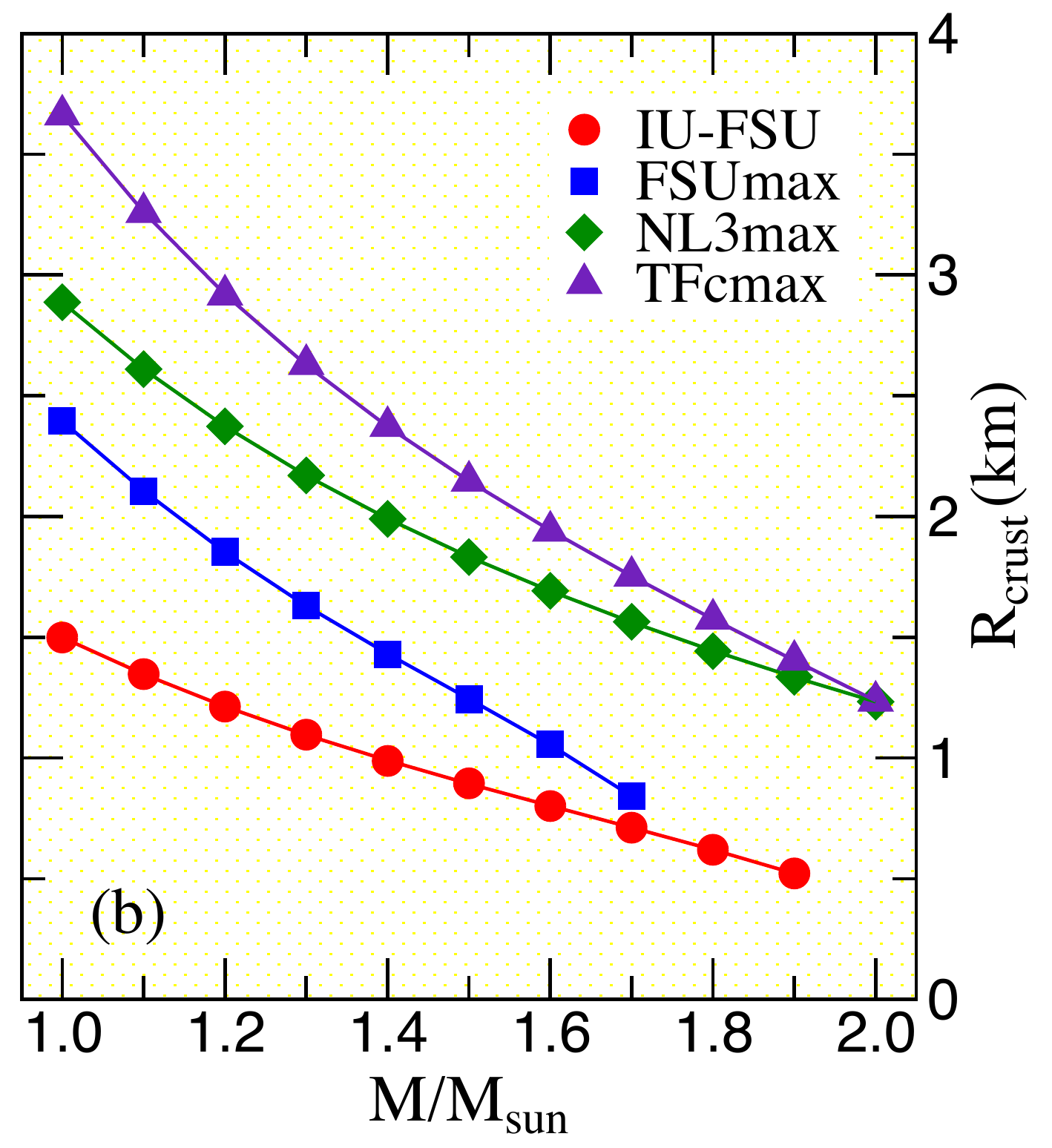}
  \caption{(Color online) Crustal contributions to the total mass (a)
  and radius (b) as a function of neutron-star mass as predicted by
  the four models considered in the text.}
 \label{Fig3}
 \end{center}
\end{figure*}

Having highlighted the most critical features of the EOS, we are now
in a position to discuss our predictions for various crustal
properties---specifically the mass, radius, and fractional moment of
inertia. We start by displaying in Fig.\,\ref{Fig3} the crustal
contribution to the mass and radius of neutron stars of various masses
as predicted by the four RMF models considered in the text. For
reference, we provide approximate analytic expressions for these two 
quantities. That is\,\cite{Fattoyev:2010tb},
\begin{subequations}
 \begin{align}
  & M_{\rm crust} \approx 8\pi \frac{R_{t}^{4}P_{t}}{R_{s}}
    \left(1\!-\!R_{s}/R_{t}\right)\,, \\
  & R_{\rm crust} \approx 8 \frac{R_{t}^{2}P_{t}}{R_{s}{\mathcal E}_{t}}
    \left(1\!-\!\frac{8R_{t}P_{t}}{R_{s}{\mathcal E}_{t}}\right)^{-1} \,.
 \end{align}
 \label{MRcr}
\end{subequations}
\begin{figure*}[htbp]
 \begin{center}
  \includegraphics[width=0.55\textwidth]{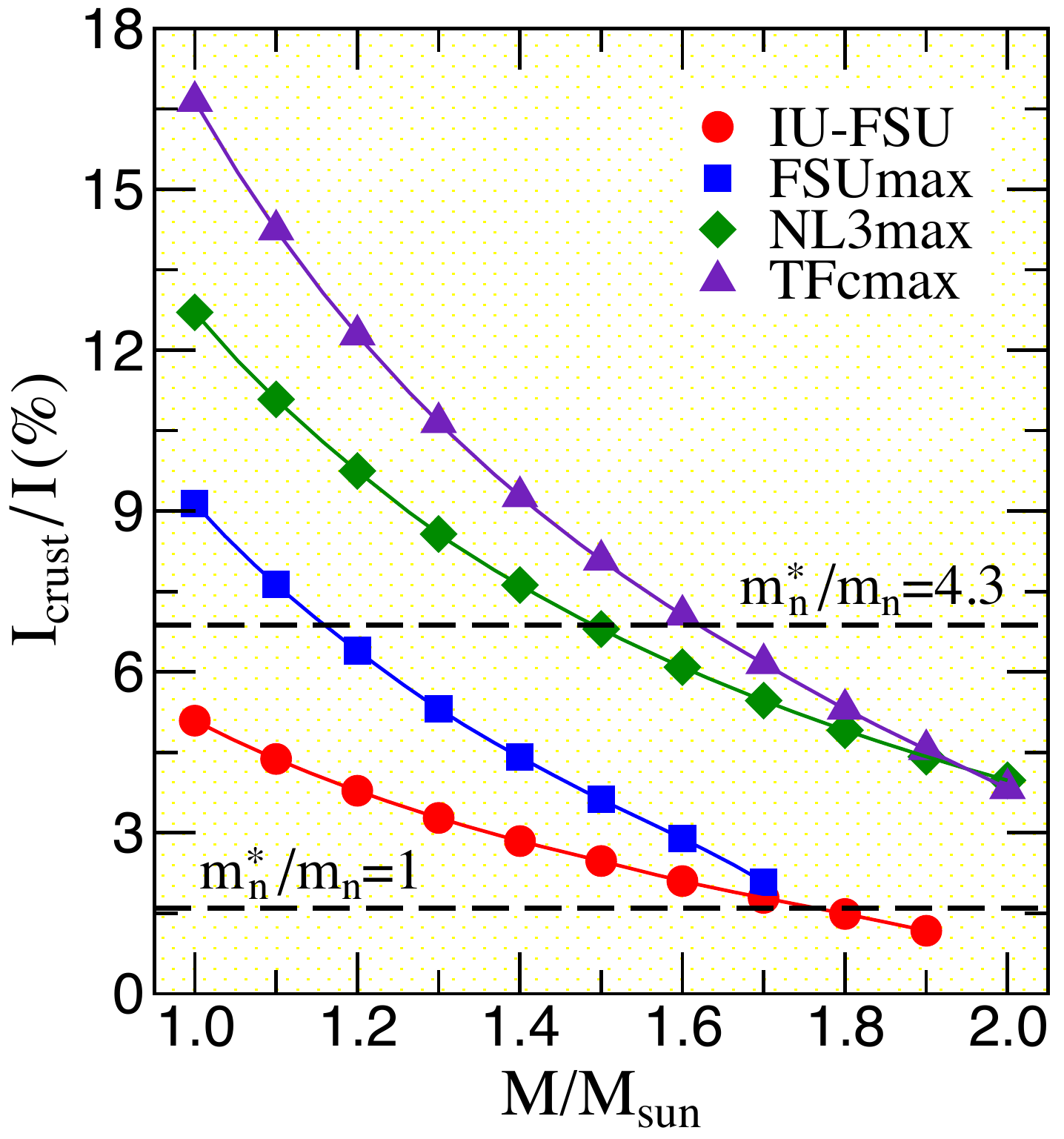}
  \caption{(Color online) Fraction of the moment of inertia residing in the crust
  as a function of stellar mass for the four models considered in the text. The
  two horizontal lines represent observational constraints from the Vela glitches
  by assuming either no ($\langle m_{n\rangle}^{\ast}/m_{n}\!=\!1$) or significant
  ($\langle m_{n}^{\ast}\rangle/m_{n}\!=\!4.3$) crustal entrainment.}
 \label{Fig4}
 \end{center}
\end{figure*}

As in the case of the crustal moment of inertia given in
Eq.(\ref{Icr1}), both $M_{\rm crust}$ and $R_{\rm crust}$ are
proportional to the product of the core radius $R_{t}$ (raised to some
power) times the transition pressure
$P_{t}$\,\cite{Fattoyev:2010tb}. Such a dependence is clearly
displayed in the figure. We observe that whereas only a very small
fraction of the total neutron-star mass resides in the low-density
crust, the crustal contribution to the overall size of the star may be
significant. For example, for the case of a 1.4\,$M_{\odot}$ neutron
star, the TFcmax model predicts that of the total 14.4\,km stellar
radius, 2.4\,km (or $\simeq$17\%) are accounted by the crust. Could
such a thick crust account for the large pulsar glitches?

To answer this question we have computed the crustal fraction of the
total moment of inertia as a function of the neutron-star mass. Our
results are displayed in Fig.\,\ref{Fig4} alongside the observational
constraints imposed from the large Vela glitches. Here we follow the
assumptions of Ref.\,\cite{Andersson:2012iu} that suggest that in
order to explain the large Vela glitches 1.6\% of the moment of
inertia must reside in the superfluid reservoir---but that such number
must be increased to almost 7\% if entrainment effects (with $\langle
m_{n}^{\ast}\rangle/m_{n}\!=\!4.3$) are included. As one may expect
from the sensitivity of crustal properties to the transition pressure,
the trends displayed in Fig.\,\ref{Fig4} follow closely those observed
in Fig.\,\ref{Fig3}. We insist that although the analytic formulas
displayed in Eqs.(\ref{Icr1}) and~(\ref{MRcr}) are highly insightful,
the predictions made for all crustal properties displayed in
Figs.\,\ref{Fig3} and~\ref{Fig4} are exact---at least within the
framework of the slow-rotation approximation. Our results indicate
that even though the observation of large pulsar glitches in the
presence of crustal entrainment provides a very stringent constraint
on the EOS, some \emph{realistic} theoretical models are consistent
with such a constraint. Indeed, our findings suggest that one may
account for the large Vela glitches even in the presence of
significant crustal entrainment provided
$M_{\rm vela}\lesssim\!1.6\,M_{\odot}$. Note that a similar conclusion
was reached in Ref.\,\cite{Steiner:2014pda} from a study of neutron-star
observations.

As mentioned earlier in Sec.\,\ref{Sec:EOS}, the complexity of the
inner stellar crust has hindered the construction of a detailed EOS and has
made us rely on a polytropic approximation that interpolates between the
solid outer crust and the uniform liquid core\,\cite{Carriere:2002bx}. Given
the importance of crustal properties on our results, it is pertinent to
examine the sensitivity of our results to our choice of $\gamma_{\rm rFG}\!=\!4/3$, 
which corresponds to the polytropic index of a non-interacting relativistic Fermi gas. 
To do so, we have examined in Table\,\ref{Table4}
the sensitivity to the choice of polytropic index on the total stellar radius,
total moment of inertia, crustal mass, crustal radius, and fractional moment
of inertia for a 1.4$M_{\odot}$ neutron star. Note that for
$\gamma_{\rm rFG}\!=\!4/3$, the results for the crustal properties are those
quoted in Table\,\ref{Table3}. Although the crustal radius displays a strong
sensitivity to the polytropic index---which correspondingly affects the total
stellar radius---neither the crustal mass nor the total and fractional moments
of inertia display dramatic changes. Indeed, Table\,\ref{Table4} displays
a change of only $\lesssim\!11\%$ for the fractional moment of inertia. This
result helps to further validate our main conclusion.

\begin{table}[htbp]
\begin{tabular}{|l||l|c|c|c|c|c|}
  \hline
  Model  & Polytrope Index & $R_{\rm tot}$(km) & $I_{\rm tot}$($10^{45}$g\,cm$^2$) &
  $M_{\rm crust}(M_{\odot})$ & $R_{\rm crust}$(km) & $I_{\rm crust}/I$(\%)\\
  \hline
  \hline
  IU-FSU & $\gamma_{\rm min}=1.000$   & 12.938 & 1.564 & 0.019 & 1.434 & 2.877\\[-0.05cm]
         & $\gamma_{\rm rFG}=1.333$   & 12.493 & 1.563 & 0.019 & 0.989 & 2.827\\[-0.05cm]
         & $\gamma_{\rm max}=2.420$   & 12.198 & 1.563 & 0.018 & 0.694 & 2.784\\
  \hline
  \hline
  FSUmax & $\gamma_{\rm min}=1.000$   & 13.647 & 1.488 & 0.028 & 2.246 & 4.559\\[-0.05cm]
         & $\gamma_{\rm rFG}=1.333$   & 12.831 & 1.487 & 0.027 & 1.430 & 4.407\\[-0.05cm]
         & $\gamma_{\rm max}=2.422$   & 12.300 & 1.486 & 0.027 & 0.900 & 4.287\\
  \hline
  \hline
  NL3max & $\gamma_{\rm min}=1.000$   & 15.610 & 1.866 & 0.054 & 3.290 & 7.970\\[-0.05cm]
         & $\gamma_{\rm rFG}=1.333$   & 14.313 & 1.862 & 0.052 & 1.990 & 7.619\\[-0.05cm]
         & $\gamma_{\rm max}=2.156$   & 13.591 & 1.860 & 0.051 & 1.267 & 7.380\\
  \hline
  \hline
  TFcmax & $\gamma_{\rm min}=1.000$   & 16.169 & 1.766 & 0.064 & 4.132 & 9.865\\[-0.05cm]
         & $\gamma_{\rm rFG}=1.333$   & 14.412 & 1.759 & 0.061 & 2.372 & 9.258\\[-0.05cm]
         & $\gamma_{\rm max}=2.007$   & 13.542 & 1.755 & 0.060 & 1.499 & 8.917\\
  \hline
\end{tabular}
 \caption{Predictions for the total radius, total moment of inertia, crustal mass, crustal radius,
 and fractional moment of inertia for a 1.4$M_{\odot}$ neutron star as predicted by the various
 models used in the text.}
 \label{Table4}
\end{table}

Before leaving this section, we would like to highlight some recent
findings on stellar radii that appear to create significant tension
with the constraints emerging from large pulsar glitches. Recently,
\citet{Guillot:2013wu} were able to determine neutron-star radii by
fitting the spectra of five quiescent low mass X-ray binaries (qLMXB)
in globular clusters. By assuming that all neutron stars independent
of their mass share a common radius ($R(M)\!\equiv\!R_{0}$) they 
determined this common radius to be
\begin{equation}
 R_{0}=9.1^{+1.3}_{-1.5}\,{\rm km}\,.
\label{RGuilliot}
\end{equation}
Although several of the assumptions made in
Ref.\,\cite{Guillot:2013wu} suggest that other interpretations may be
favored\,\cite{Lattimer:2013hma}, Eq.\,(\ref{RGuilliot}) is highly
intriguing as it suggests stellar radii significantly smaller than
those predicted by many theoretical
models\,\cite{Lattimer:2006xb,Horowitz:2014bja}. Moreover, as one
combines such a small radius with the observation of neutron stars
with masses of $M\!\approx\!2
M_{\odot}$\,\cite{Demorest:2010bx,Antoniadis:2013pzd}, then
\emph{most} theoretical models appear to be ruled out. This is
because, on the one hand, the pressure around twice nuclear-matter
saturation density must be small enough to accommodate small stellar
radii but, on the other hand, the pressure must stiffen significantly
at high densities to support massive neutron stars.

\begin{figure*}[htbp]
 \begin{center}
  \includegraphics[width=0.65\textwidth]{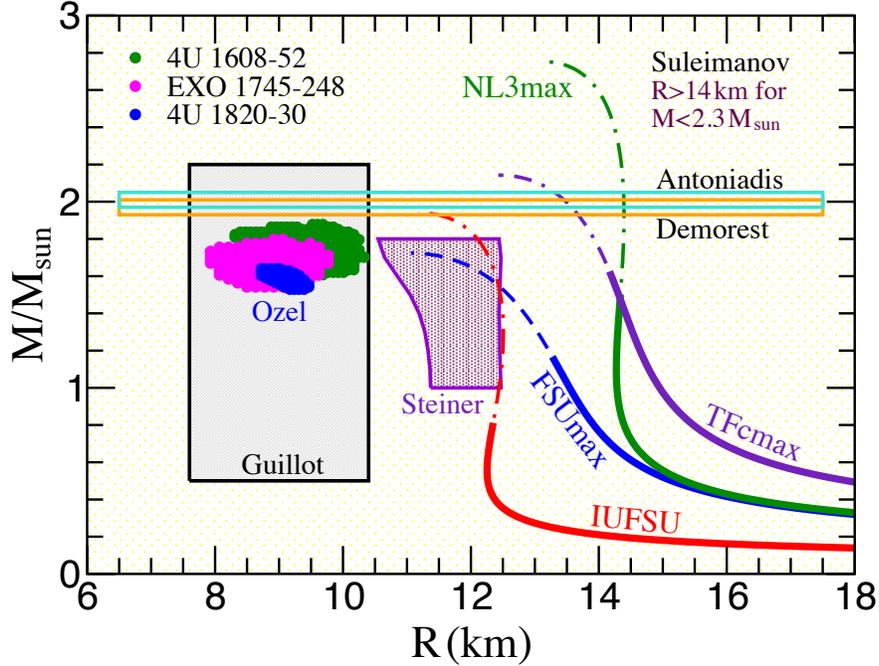}
    \caption{(Color online) \emph{Mass-vs-Radius} relationship predicted by the four models
    considered in the text. Also shown are observational constraints on stellar
    masses\,\cite{Demorest:2010bx,Antoniadis:2013pzd} and
    radii\,\cite{Ozel:2010fw,Steiner:2010fz,Suleimanov:2010th,Guillot:2013wu}, as explained
    in the text.}
 \label{Fig5}
 \end{center}
\end{figure*}

Such a situation is illustrated in Fig.\,\ref{Fig5} where the
\emph{Mass-vs-Radius} relationship is displayed for the four models
considered in the text. The solid portion of each of these curves
indicates the region that is consistent with the constraint from the
large Vela glitches. That is, whereas IUFSU is compatible only with
very small masses, TFcmax is
consistent up to a mass of $1.6\,M_{\odot}$. In addition to the
``single-radius'' finding of \citet{Guillot:2013wu}, we display
results obtained from the analysis of several X-ray bursters
by\,\citet{Ozel:2010fw,Steiner:2010fz}, and
\citet{Suleimanov:2010th}. However, given that systematic
uncertainties in the analysis of X-ray bursters continue to hinder the
reliable extraction of stellar radii\,\cite{Suleimanov:2010th}, qLMXBs
may provide, at least at present, a more reliable source for the
determination of stellar radii. Also shown in the figure are the
limits imposed by the measurement of two massive neutron
stars\,\cite{Demorest:2010bx,Antoniadis:2013pzd}. It is evident from
Fig.\,\ref{Fig5} that none of our theoretical models are consistent
with stellar radii significantly smaller than 12\,km. Indeed, to our
knowledge very few---perhaps only one\,\cite{Wiringa:1988tp}---EOS can
simultaneously account for small radii and large masses. Note that we
are unaware on whether such an EOS may be able to account for the
large Vela glitches.  In our case models such as TFcmax, that appear
to account for both massive neutron stars and large pulsar glitches,
predict stellar radii that are significantly larger than the common
radius value of $R_{0}\!=\!9.1^{+1.3}_{-1.5}\,{\rm km}$. Indeed, for a
$1.4\,M_{\odot}$ neutron star, TFcmax predicts a radius of
14.4\,km---or 4\,km larger than the suggested upper limit for~$R_{0}$.

\section{Conclusions}
\label{Conclusions}
The large jumps observed in the spin frequency of some neutron stars
like the Vela pulsar suggest the existence of an angular-momentum
reservoir confined to the inner stellar crust. As such, large pulsar
glitches have been used to constrain the EOS of neutron-rich matter by
imposing limits on the fraction of the moment of inertia that must
reside in the crust. Until recently, the large Vela glitches demanded
that at least 1.6\% of the stellar moment of inertia must be located
in the crust. Although useful, many realistic EOS are compatible with
such limit. However, recent studies seem to indicate that a
significantly larger value is required once crustal entrainment is
taken into account. Indeed, encoding the impact of crustal entrainment
in a value of the effective neutron mass of $\langle
m_{n}^{\ast}\rangle/m_{n}\!=\!4.3$, suggests that the limit on the
fractional moment of inertia must be increased to almost 7\%. Unless
the mass of the Vela pulsar is very small ($\lesssim\!1\,M_{\odot}$)
then a large number of theoretical models become incompatible with
this much more stringent limit. This has lead to the assertion that
the ``crust is not enough".

The main goal of this work was to re-examine whether indeed the crust
is not enough. Given that the crustal moment of inertia is highly
sensitive to the transition pressure $P_{t}$ at the crust-core
interface, we examined the predictions of a variety of relativistic
mean-field models in the transition region.  In particular, we found
that certain bulk properties of the EOS at the crust-core interface
are strongly correlated to the slope of the symmetry energy $L$.
However, not the transition pressure. Indeed, $P_{t}$ increases
monotonically for small $L$, reaches its maximum at some intermediate
value, and ultimately decreases with increasing $L$. Given that in the
class of RMF models used in this work one can easily tune the value of
$L$, we searched for models with the largest transition pressure. By
doing so, we were able to generate neutron stars with thick crusts and
large crustal moments of inertia. Indeed, our results support the
standard model of pulsar glitches---provided the Vela mass does not
exceed $1.6\,M_{\odot}$. In particular, this requires values for the
neutron-skin thickness of ${}^{208}$Pb of about 
$R_{\rm skin}^{208}\!\simeq\!(0.20\!-\!0.26)\,{\rm fm}$. This finding---which
represents the central result of our work---offers yet another
attractive connection between a laboratory measurement and an
astrophysical observation.

We close with a few questions and comments on the impact of our findings
on other neutron-star observables sensitive to crustal dynamics. First,
rapidly rotating neutron stars with large asymmetries (such as mass
quadrupoles) are efficient sources of gravitational waves. In particular,
``large mountains'' on rapidly rotating neutron stars may efficiently radiate
gravitational waves provided that the breaking strain of the crust is
large\,\cite{Andersson:2009yt}. Horowitz and Kadau have performed
large-scale Molecular-Dynamics simulations that reveal a large breaking
strain in support of large mountains in neutron stars\,\cite{Horowitz:2009ya}.
Such findings suggest that large mountains in rapidly rotating neutron stars
may generate gravitational waves that may be detectable by the next generation
of gravitational-wave observatories. Thus, one would like to understand
the impact of the large crustal thicknesses found in our work
({\sl e.g.,} $R_{\rm crust}\!\simeq\!2.4\,{\rm km}$ for a $1.4\,M_{\odot}$
neutron star) on the breaking strain of the crust and ultimately on the
emission of gravitational waves.
Moreover, the size of the crust may also impact the
cooling light curves of low-mass X-ray binaries during quiescence.
Indeed, a thick stellar crust could increase the time scale for crustal
cooling after extended periods of accretion; see
Ref.\,\cite{Brown:2009kw} and references contained therein.

However, although the large crustal thicknesses raises interesting
questions and accounts for large crustal fractions of the stellar
moment of inertia, the moderate values of $L$ required to account for
the large pulsar glitches are inconsistent with a recent analysis of
quiescent low mass X-ray binaries that suggests very small stellar
radii. In fact, the class of relativistic mean-field models employed
in this work are unable to generate such small radii---regardless of
whether one incorporates the pulsar-glitch constraint or not. Although
not without controversy, we found instructive to take the small-radius
result at face value and ask whether it is possible to account for
both large pulsar glitches and small neutron stars. If crustal
entrainment is as large as it has been suggested, then a moderate
value of the neutron-skin thickness of ${}^{208}$Pb is required. This
prediction can be tested by performing parity violating electron
scattering experiments at the Jefferson Laboratory.  However, in order
for neutron-star radii to be as small as suggested, a dramatic
softening of the symmetry energy must develop by the time that the
density reaches about 2 times nuclear-matter saturation density. Such
a rapid softening is likely to involve a change in the structure of
dense matter---likely due to a phase transition. In principle, the
onset of such a phase transition may also be probed in the laboratory
using energetic heavy-ion collisions of highly-asymmetric
nuclei. Finally, the EOS must significantly stiffen at even higher
densities in order to account for the observation of massive neutron
stars.  Such unique behavior of the EOS and its possible realization
in neutron stars reaffirms the special bond between astrophysics and
nuclear physics.


\begin{acknowledgments}
\vspace{-0.4cm} This work was supported in part by grants
DE-FG02-87ER40365 and DE-FD05-92ER40750 from the U.S. Department of
Energy, by the National Aeronautics and Space Administration under
grant NNX11AC41G issued through the Science Mission Directorate, and
the National Science Foundation under grant PHY-1068022.
\end{acknowledgments}
\vfill

\bibliography{Glitches.bbl}
\vfill\eject
\end{document}